\documentclass[12pt,preprint]{aastex}

\slugcomment{Accepted for publication in ApJ}

\begin{document}

\shorttitle{Critical rotation and quiescence in collapsars}

\shortauthors{L\'{o}pez-C\'{a}mara, Lee \& Ramirez-Ruiz}

\title{Critical angular momentum distributions in collapsars:
  quiescent periods from accretion state transitions in long gamma-ray
  bursts}

\author{Diego L\'{o}pez-C\'{a}mara\altaffilmark{1,2}, William
  H. Lee\altaffilmark{2} and Enrico
  Ramirez-Ruiz\altaffilmark{3}}\altaffiltext{1}{Instituto de Ciencias
  Nucleares, UNAM, Apdo. Postal 70-543, M\'{e}xico D.F. 04510,
  MEXICO}\altaffiltext{2}{Instituto de Astronom\'{\i}a, UNAM,
  Apdo. Postal 70-264, M\'{e}xico D.F. 04510,
  MEXICO}\altaffiltext{3}{Department of Astronomy and Astrophysics,
  University of California, Santa Cruz, CA 95064, USA}

\begin{abstract}

The rotation rate in pre-supernova cores is an important ingredient
which can profoundly affect the post-collapse evolution and associated
energy release in supernovae and long gamma ray bursts
(LGRBs). Previous work has focused on whether the specific angular
momentum is above or below the critical value required for the
creation of a centrifugally supported disk around a black hole. Here,
we explore the effect of the {\em distribution} of angular momentum
with radius in the star, and show that qualitative transitions between
high and low angular momentum flow, corresponding to high and low
luminosity accretion states, can effectively be reflected in the
energy output, leading to variability and the
possibility of quiescent times in LGRBs.
\end{abstract}


\section{Introduction}\label{sec:intro}

The collapse of massive stellar cores producing Supernovae (SN) has
been clearly linked to the production of Long Gamma Ray Bursts (LGRBs)
in the last ten years (see \citet{wb06,grrf09} for reviews), providing
tantalizing clues to the progenitors and the environments in which
they occur \citep{fruchter06}, as well as providing fresh glimpses of
the high-redshift universe \citep[e.g.,][]{prochaska09}. A key ingredient in
this context is the stellar rotation rate, impacting the energy
release and general outcome of the event following the implosion of
the Chandrasekhar-mass iron core. Stellar evolution considerations
have shown that it is non-trivial to have rapid rotation in the inner
regions of the star, as mass loss and magnetic fields both conspire to
reduce the pre-SN angular momentum. Calculations
\citep[hereafter WH06]{spruit02,hws05,yoon05,wh06} indicate that the
final distributions of specific angular momentum may not easily lead
to the production of a centrifugally supported accretion disk
following collapse, with the envelope experiencing essentially radial
infall. This has generally been believed to be insufficient to power
LGRBs since the lack of shocks and the associated dissipation are
unable to transform gravitational binding energy into radiation
effectively. A possible evolutionary channel that avoids some of the
pitfalls associated with the giant phase was proposed in WH06 and by
\citet{yoon05}, in which thorough mixing on the main sequence in very
massive stars reduces mass and angular momentum losses. However, a
large class of stars remains in which rotation will be substantially
slowed in late evolutionary stages, thus influencing the outcome of
the collapse.

The low angular momentum regime in collapsars has not been fully
explored. In most cases, studies of the attending neutrino cooled
accretion flows have considered specific rotation laws that guarantee
by a large margin the formation of a centrifugally supported disk,
either because the angular velocity is assumed to be nearly Keplerian,
or because the absolute value of the angular momentum given implies a
circularization radius much larger than the radius of the innermost
stable circular orbit (ISCO), which in General Relativity is $r_{\rm
  ISCO}=3 r_{\rm g}$ for a Schwarzschild black hole, where $M_{\rm BH}$ is the black hole mass and
$r_{\rm g}=2GM_{\rm BH}/c^{2}$
\citep{mw99,pwf99,h00,npk01,pb03,pmab03,lrrp05,fujimoto06,nagataki07}. We
previously studied \citep[hereafter Paper~I]{lrr06,lclrr09}, cases at
the threshold for disk formation, where a centrifugally supported disk
gives way to nearly radial inflow due to relativistic effects, finding that even in the case of
slow rotation, enough energy may be available through the formation of
{\em dwarf} disks \citep{ib01,zb09} in near free fall to power
LGRBs. However, the distributions of specific angular momentum we
considered were constant in the equatorial plane, and smoothly
decreasing towards the rotation axis. This is unrealistic, as the
specific angular momentum {\em increases} outwards in the core and
envelope, with marked transitions at the boundaries between different
layers in the star (see, e.g., Figure~2 in WH06).

Further, as the newborn black hole (BH) accretes infalling matter,
both its mass and angular momentum increase, raising the threshold
value of the critical angular momentum required for the formation of a
centrifugally supported disk, $J_{\rm crit}=2 r_{\rm g} c$. The competition with the
increase of angular momentum in the infalling material determines the
large scale properties of the flow. This was recently pointed out and
addressed by \citet{jp08} in simple form, generically by
\citet{knj08a,knj08b} in studying the effect of the distribution of
angular momentum in the star on the light curve of LGRBs, and more
recently by \citet{lindner09} in two-dimensional simulations
of collapsar accretion.

In this study, we explore how different distributions of angular
momentum as a function of radius, can have important effects on the
qualitative properties of the accretion flow, and hence on the accretion rate and global energy release, which we quantify through the neutrino luminosity, $L_{\nu}$. We pay particular attention to the general form and rate of
increase of specific angular momentum with radius in the star in this
respect, and show that state transitions may in principle produce
observable consequences in LGRBs relevant to variability and periods
of quiescence. In section \S~\ref{sec:input} we describe the
assumptions and approximations made in our calculations and identify
the numerical setup. Section \S~\ref{sec:results} is devoted to presenting the
flow transitions which result from different angular momentum
distributions, and \S~\ref{sec:states} to the conditions necessary for
them to occur. In \S~\ref{sec:critical} we combine these, given
realistic angular momentum distributions, to investigate episodic
energy release from accretion. Our conclusions and prospects for
observability are discussed in \S~\ref{sec:discussion}.
      
\section{Setup and physics}\label{sec:input}

The evolution of the infalling core and envelope after the formation
of the central black hole from the stellar Iron core is followed with
the same azimuthally symmetric, two dimensional Smooth Particle
Hydrodynamics (SPH) code as in Paper~I. The computational domain
covers the region between spherical radii $R_{\rm in}=2 \times
10^{6}$~cm and $R_{\rm out}=2 \times 10^{9}$~cm, without assuming
reflection symmetry with respect to the equatorial plane. The
equation of state is contains contributions to the total pressure $P$
from an ideal gas of $\alpha$ particles and free nucleons in nuclear
statistical equilibrium (NSE), $P_{\rm gas}$, black-body radiation,
$P_{\rm rad}$, relativistic e$^{\pm}$ pairs of arbitrary degeneracy,
$P_{\rm e^{\pm}}$ \citep{bdn96} and neutrinos. Neutronization, and a
variable electron fraction are computed assuming charge neutrality and
weak equilibrium \citep{bel03,lrrp05}, depending on whether the fluid
is optically thin or thick to neutrinos. The neutrino emissivities
which dominate the cooling arise from $e^{\pm}$ capture onto free
nucleons and e$^{\pm}$ annihilation, and are computed from the
tabulated results of \citet{lmp01} and the fitting functions of
\citet{i96}, respectively. A two-stream approximation including these
processes, and coherent scattering off free nucleons was implemented
to compute the local cooling rate \citep{pn95,dmpn02,jypdm07}. The neutrino luminosity $L_{\nu} $ is computed as the volume integral of the local neutrino emissivity, given the thermodynamical conditions present (see eq.8 in Paper~I). The equations of motion also include a full expression for the viscous
stress tensor $t_{ij}$ in azimuthal symmetry. We compute the
coefficient of viscosity with the standard $\alpha$ prescription of
\citet{ss73}, namely $\eta_{\rm v}=\alpha \rho c_{\rm s} H$, where
$c_{\rm s}$ is the local sound speed, $H=c_{\rm s} \Omega$ is the
pressure scale height, and $\Omega$ is the local Keplerian orbital
frequency. We have used $\alpha=0.1$ throughout for the simulations
described in this paper. Given that this form is only applicable if
there is a rotating accretion structure, in the usual sense of largely
centrifugal support with a additional pressure corrections, we use a
switch (similar to that described by \citet{mw99}) so the viscosity
only operates when rotation dominates over essentially free infall. In
practice this is carried out by comparing the local radial and
azimuthal components of velocity, and allowing $\eta_{\rm v}\neq 0$
only if $v_{r}/v_{\phi} \leq 1$ (the functional form is made continous
to avoid spurious transitional behavior when rotational support
dominates).

As a generic initial condition we consider the 1D pre-supernova models
of WH06. Specifically we used model 16TI, a rapidly rotating ($v_{\rm
  rot}=390$km~s$^{-1}$), low metallicity ($Z=0.01Z_{\odot}$) WR star
of 16M$_{\odot}$ with low mass loss ($2M_{\odot}$ total loss up to the
pre-supernova stage), with an Iron core of $M_{Fe}=1.6M_{\odot}$. This
particular model did not consider the effects of internal magnetic
fields. The density, temperature and radial velocity of the stellar
core was mapped to two dimensions assuming spherical symmetry, while
the Iron core was condensed onto a point mass at the origin
representing the newly formed black hole and producing a
pseudo-Newtonian potential according to the expression of
\citet{pw80}, $\phi=-GM_{\rm BH}/(R-r_{\rm g})$. This reproduces the
existence and location of the innermost stable circular orbit (ISCO)
of the Schwarschild solution in General Relativity, which is the
leading effect in terms of separating the potential progenitors of
GRBs into those which can eventually form a centrifugally supported
accretion disk and those that cannot.

The distributions of specific angular momentum, $\mathcal J$ in
pre-supernova stars, as computed in evolutionary codes, generally
follow a monotonic increase with radius, with sharp transitions at the
interfaces between shells where the composition changes
abruptly. Thus, to study the collapse, we separated the angular
momentum distribution into radial and polar angle components as
$\mathcal J=\mathcal J(R,\theta)=J(R) \Theta(\theta)$, where $R$ is
the spherical radius and $\theta$ is the polar angle, and assumed
rigid body rotation on shells, with $\Theta(\theta)=
\sin^{2}\theta$. For the radial component, we considered various
functional forms: a) linearly increasing with radius, $J(R) \propto
R$; b) constant, $J(R)=J_{0}$ with one or more superimposed sharp
increases, or spikes to mimic the transitions at shell boundaries; c)
the distribution given in WH06 for model 16TI, multiplied by a
normalization factor of order unity.

\section{Flow transitions}\label{sec:results}

\subsection{Linear distributions of specific angular momentum.}\label{sec:linJ}

In Paper~I, we considered simple, constant distributions of $J(R)$ in
order to gauge the effect of the absolute value of angular momentum
and the general properties of the flow when compared to previous work,
and provide a further guide. When $J(R) < J_{\rm crit}$ no
centrifugally supported disk formed, and as the black hole accreted
matter and increased its mass, so did the value of $J_{\rm crit}$,
further inhibiting the creation of a disk. The energy released came
from the equatorial compression of the infalling gas, with peak neutrino
luminosities $L_{\nu}\simeq 10^{51}$~erg~s$^{-1}$. On the other hand,
when $J(R) \ge J_{\rm crit}$, a hot shocked torus promptly formed, but
the mass accretion rate onto the black hole, $\dot{M}_{\rm
  BH}\sim$0.1-0.5$M_{\odot}$~s$^{-1}$, was such that the equatorial
angular momentum in the flow was always above $J_{\rm crit}$, and the
disk was never destroyed.

Given the general rising trend in the distribution of specific angular
momentum in pre-SN models mentioned above, we first considered linearly
increasing distributions in $J(R)$ by writing $J(R) = J_{0} + m (R /
R_{\rm out})$ and computing cases with varying $J_{0}$ and $m$. For
low ($J_{0} \ll J_{\rm crit}$) or high ($J_{0} \gg J_{\rm crit}$)
values of the starting point in the distribution, and independently of
the value of $m$, the result was akin to that previously obtained,
namely, a quasi-radial inflow (QRI) or a long-lived accretion disk
around the BH, respectively.

\begin{figure}[h!]
  \begin{center}
  \epsscale{1.0}    
   \plotone{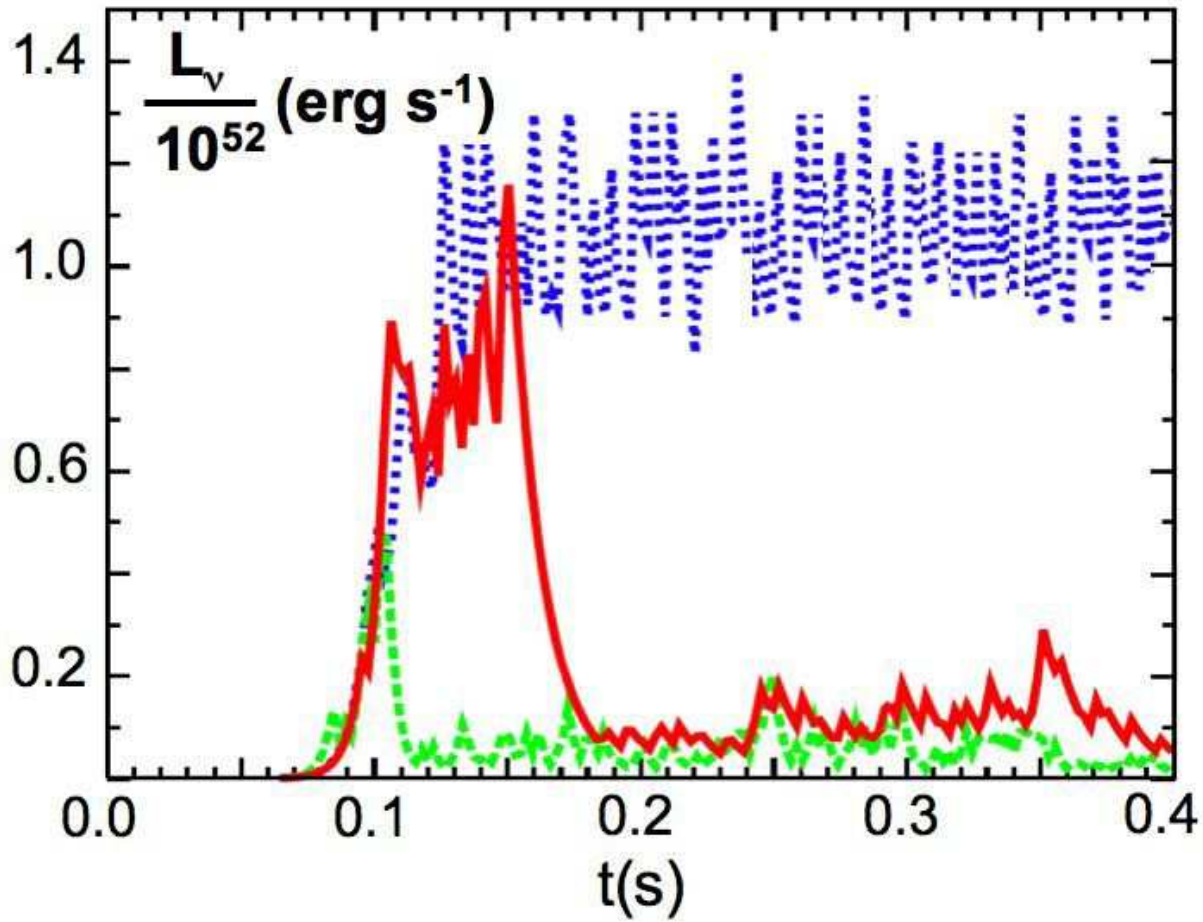} 
    \caption{Neutrino luminosities (scaled to $10^{52}$~erg~s$^{-1}$)
      for simulations with initially linear distributions of specific
      angular momentum, $J(R)=J_{0}+m R/R_{\rm out}$. The curves show
      the cases with $J_{0} / J_{\rm crit} = 1.05$, for which the
      angular momentum increased rapidly (blue, $m=3$, persistent
      disk); slowly (green, $m=1$, no disk); and moderately (red,
      transitory disk correspondent to the critical value $m=2$).}
    \label{jlin}
  \end{center}
\end{figure}

Intermediate cases, however, here when $ 0.85 \le J_{0} / J_{\rm crit}
\le 1.15$, exhibit diverse behavior depending on the rate of increase
in $J(R)$, given by $m$. Setting $J_{0} / J_{\rm crit} = 1.05$, for
$m\simeq 2$, a centrifugally supported torus momentarily appears as
the gas encounters the centrifugal barrier, and is subsequently
accreted after a delay of $\simeq 0.1$~s. This is simply a
manifestation of the well-known ``runaway radial'' instability
\citep{acn83}, triggered by the rapid rise in black hole mass and thus
$J_{\rm crit}$. Moreover, the short-lived disk is not able to perturb
the upstream gas since the inflow is supersonic downstream of the shock
front. Once the critical rotation rate rises above that of the angular
momentum in the envelope, the gas is accreted essentially in a free
fall time scale and the neutrino luminosity drops by about one order
of magnitude, to $L_{\nu} \simeq 10^{51}$~erg~s$^{-1}$. There is thus
no ``memory'' in the flow, in the sense that once the transitory disk
is destroyed, the subsequent material evolves as if the disk had never
appeared. Figure~\ref{jlin} shows the neutrino luminosity $L_{\nu}$
for the cases where the angular momentum increases either rapidly
(blue line), slowly (green line), or moderately (red solid line).

It is thus clear, as explored initially by \citet{jp08}, that the
competition between the rise in angular momentum as one moves outward
in the star, and the increase in BH mass, plays a fundamental role in
the establishment of an accretion structure capable of releasing a
significant amount of energy deep in the gravitational potential
well. Furthermore, even a smooth, monotonically increasing
distribution of $J(R)$, can produce temporal transitions between
qualitatively different accretion states, namely, a QRI and a centrifugally supported disk.

\subsection{Rapid transitions in the angular momentum distribution}
\label{sec:spikes}

The particular feature of strong jumps in $J(R)$ between neighboring
shells found in stellar evolution calculations led us next to explore
how such a qualitative variation in the angular momentum distribution
can alter the properties of the accretion flow. For this, we
considered a constant background distribution, with $J_{0} \lesssim
J_{\rm crit}$, over which two radial shells where $J(R)$ reaches $1.5
\ J_{\rm crit}$ are overlaid. Their width is $dR \simeq
5\times10^{4}$~cm, and they are separated by $\approx 2~dR$.

\begin{figure}[h!]
  \begin{center}
  \epsscale{1.9}    
   \plotone{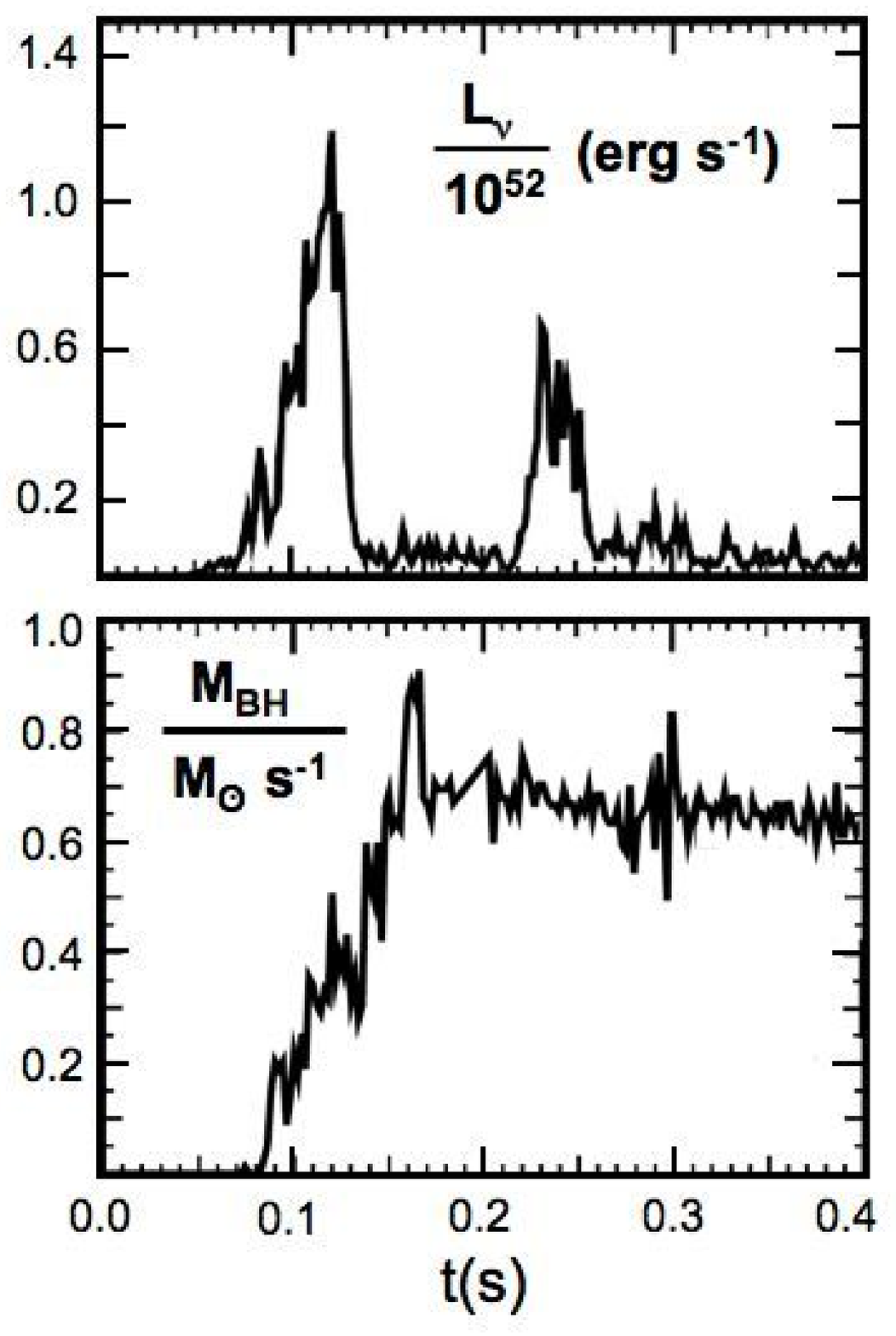} 
    \caption{Neutrino luminosity (top) and black hole mass accretion
      rate (bottom) for a constant distribution of specific angular
      momentum with $J(R)<J_{\rm crit}$, over which two narrow radial
      spikes with $J(R)>J_{\rm crit}$ have been overlaid. The
      luminosity and black hole mass accretion rate rise and fall with
      the arrival and destruction of each rapidly rotating shell at
      the centrifugal barrier.}
    \label{spike}
  \end{center}
\end{figure}

Since the angular momentum of the material initially reaching the BH
is below the threshold for disk formation, a QRI promptly forms. Once
the fluid in the rapidly rotating shells reaches the centrifugal
barrier, this gives way to a shocked accretion disk, and both the
accretion rate and neutrino luminosity rapidly rise (see
Figure~\ref{spike}). The delay within which the disk forms is
essentially the free fall time of the inner shell, $t_{\rm ff} \propto
\overline{\rho}^{\ -1/2} \sim 0.1$~s. In this simulated case, the
inter-shell region contains enough material to completely overwhelm
the first disk, leading to a second episode of QRI. This now lasts until the arrival of the second shell, with a
delay again given by its proper free fall time. Even though the
transitions are clearly visible in both the accretion rate and the
luminosity, some differences clearly stand out. First, while
$\dot{M}_{\rm BH}$ decreases by less than 30\%, the luminosity drops
by more than one order of magnitude. The discrepancy is due to the
fundamental morphological change in the accretion flow: the QRI has a
high mass accretion rate but an extremely low efficiency for
converting gravitational potential energy into internal energy (it
resembles Bondi accretion in this sense), while the centrifugally supported
accretion disk allows for substantial energy conversion and
dissipation, leading to higher luminosities. The associated accretion
efficiencies through neutrinos, $\eta_{\nu, \rm acc}=L_{\nu}/\dot{M}_{\rm BH}c^{2}$ are in the expected range of (1-10)\% for differentially rotating neutrino
cooled disks \citep{dmpn02,lrrp05}, and a factor of 10 or more below
this for the QRI, where the emission is due to equatorial compression of the flow to a degree which depends on the value of the angular momentum used \citep[Paper I]{lrr06}. Second, there is a slight delay in the correlated temporal variations of $L_{\nu}$ and $\dot{M}_{\rm BH}$. Those in the former come slightly earlier, since the spikes in the accretion rate actually correspond to the mass accreted as the slowly rotating shells approach the black hole and the differentially rotating disk is destroyed.

The duration of each interval is dictated in this case by the initial
radial position and extent of each shell and the corresponding free
fall times. Since both shells here have the same radial extent and
their initial radii are comparable, the associated intervals of
activity are comparable. Variations in the 
form and normalization of the rotation rate as a function of radius
thus impinge upon the duration of active periods and their relative
power output.

We thus conclude that the presence of rapidly rotating shells in a
slowly rotating background flow can lead to transitions between the
QRI, low-luminosity and the disk, high-luminosity states, with delays
that are correlated with the initial position and width of such
shells. This is valid as long as the upper and lower limits in the
distribution lie below and above the critical rotation rate for the
existence of a centrifugal barrier. With this in mind, we can now
address possible effects in GRB central engines in more realistic way,
using the particular distributions found in WH06.

\section{State transitions and evolutionary pathways}
\label{sec:states}

Before we directly analyze the behavior resulting from the
evolutionary model of WH06, let us consider under which conditions an
infalling shell can destroy a centrifugally supported disk.

Qualitatively, one would expect that the relative mass of successive
collapsing regions should play a role in determining the outcome of
the flow structure. The limits are easy to visualize, as a low mass
shell with little or no rotation will likely only perturb an existing,
massive centrifugally supported disk. Conversely, a massive flow in
near free fall, due to its high ram pressure, will destroy the small
pre-existing disk. So where does the transition between the two types
of flow lie? To address this issue, we performed generic calculations
(again with model 16TI), and included variations in angular momentum,
so that the masses of successive shells with high ($J_{\rm
  high}=3r_{\rm g} c$, leading to a first disk) and low ($J_{\rm low}=1.9
r_{\rm g} c$, leading to QRI) angular momentum were $M_{\rm d}$ and
$M_{\rm QRI}$, respectively. Their position and width were adjusted to explore a
range in $\mu=M_{\rm d}/M_{\rm QRI}$ from 0.1-10. An outer shell with
high rotation will lead invariably to the formation of a centrifugally
supported disk at late times. Note that since we are always assuming
rigid body rotation on shells through $\Theta(\theta)$, it is only the
ratio of the masses that is relevant here.

\begin{figure}[h!]
  \begin{center}
   \epsscale{0.7}
   \plotone{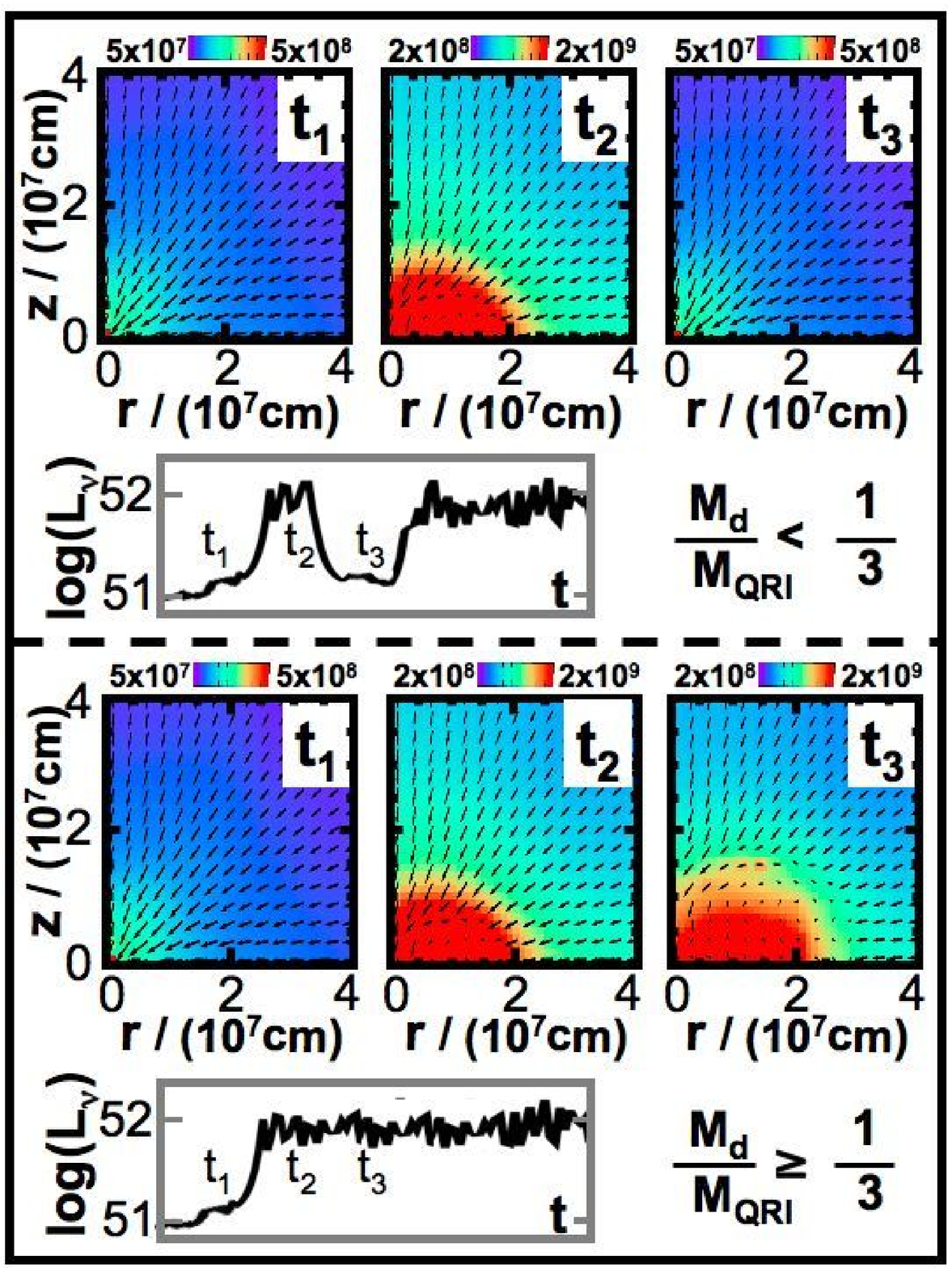} 
    \caption{Density (in g~cm$^{-3}$) and velocity field showing the
      flow morphology during the collapse in the stellar core. The
      snapshots, at $t_1 \leq t_2 \leq t_3$, are taken from
      simulations of model 16TI of WH06 for varying normalization
      factors in the radial distribution of specific angular
      momentum. Two possible evolutionary pathways are indicated. (a)
      Top: When $\mu=M_{\rm d}/M_{\rm QRI} < 1/3$, the initial QRI
      ($t_1$) is followed by the formation ($t_2$), destruction
      ($t_3$) and re-emergence of a centrifugal accretion disk. (b)
      Bottom: When $\mu \geq 1/3$, once a disk of mass $M_{\rm d}$
      forms, it persists throughout the evolution as a coherent
      structure, surviving the infall of a low-angular momentum shell
      of mass $M_{\rm QRI}$ unperturbed. The transitions are
      reflected in the neutrino luminosity.}
        \label{fig:pathways}
  \end{center}
\end{figure}

For the parameters given above, we find that if $\mu \leq 1/3$, the
low-angular momentum QRI fully destroys the pre-existing accretion
disk on a dynamical time scale. Later, when the second shell with
$J=J_{\rm high}$ reaches the centrifugal barrier, a new disk is
created, which persists as long as the inflow has sufficient
rotation. This is illustrated in the upper panel of
Figure~\ref{fig:pathways}, where the velocity field shows the flow
morphology, and in particular the formation of shocks in the disk at
various stages of the collapse. For this regime, after an initial QRI
($t=t_1$) before the initial shell with $J=J_{\rm high}$ reaches the
inner boundary, a centrifugally supported disk is created ($t=t_2$)
and subsequently destroyed ($t=t_3$). The associated neutrino
luminosity shown in the top panel of Figure~\ref{fig:pathways}
exhibits a high-state stage between two low-state
episodes.

For $\mu \geq 1/3$, the initial disk is able to absorb the impact of
the infalling shell and survive as a coherent structure. Thus, when
the second shell with rapid rotation approaches it merely adds to the
existing disk activity, and continues its evolution rather unperturbed
(bottom panel in Figure~\ref{fig:pathways}). The light curve in this
case shows continuous emission at the high state for $t>t_{2}$.

We note here that the absolute values of $J_{\rm high}$ and $J_{\rm
  low}$ represent an additional factor which can alter the resulting
state transitions. Morphologically, the QRI resembles Bondi accretion,
where technically $J_{\rm low}=0$, and it is certainly inefficient in
terms of energy release. But even a rotation rate below the critical
value can have non-negligible consequences when combined with a shell
slightly above this threshold, and alter the determination of the
critical mass ratio $\mu=M_{\rm d}/M_{\rm QRI}$ required for the
transition to occur. In an attempt to quantify this effect, we carried out several additional simulations. In the first case, we used $\mu=1/3$ and $J_{\rm high}=3r_{\rm g}c$ as before, but set $J_{\rm low}=0$, thus forcing the more massive shell into strict radial infall. The interaction now resulted in the full destruction of the centrifugally supported accretion disk. In the second, we changed the mass ratio to $\mu=1$ and used $J_{\rm high}=3r_{\rm g}c$ and $J_{\rm low}=0$. This time, the centrifugally supported disk survived during the entire simulation. 

A simple explanation of this trend can be given as follows. The critical angular momentum for the appearance of the centrifugal barrier is $J_{\rm crit}=2 r_{\rm g}c$, and we write $J_{\rm low}=A r_{\rm g}c$, where $A$ is a normalization parameter. Keeping $J_{\rm high}=3 r_{\rm g} c$ as in the above examples, and assuming the result after both shells have approached the black hole is a mixed configuration where the available angular momentum has been distributed into both components, the specific angular momentum is now $J^{\prime}$. In order for the disk to survive, we must have $J^{\prime} \geq J_{\rm crit}$, which translates after a few lines of algebra into a condition on the masses as $M_{\rm d} \geq (2-A) M_{\rm QRI}$, or $\mu \geq (2-A)$. For $A=1.9$, as used in our standard case, this implies $\mu \geq 1/10$, roughly consistent with the factor $\mu=1/3$ found above. If $A=0$, as now tested, this yields $ \mu \geq 2 $ as a condition for disk survival, i.e., it must be more massive in order to avoid destruction. In the first test described above, where $J_{\rm low}$ is drastically reduced, the outcome is as expected and the disk disappears. In the second, the rise in disk mass is able to offset the drop in $J_{\rm low}$ and the centrifugally supported disk persists. The fact that the normalization in mass does not strictly follow our simple analytical prescription is clearly due to an oversimplification of the dynamics, but it appears nonetheless to capture the essential aspect of the problem. 

\section{Critical distributions of angular momentum and episodic energy release}\label{sec:critical}

\begin{figure}[h!]
  \begin{center}
   \epsscale{1.0}  
   \plotone{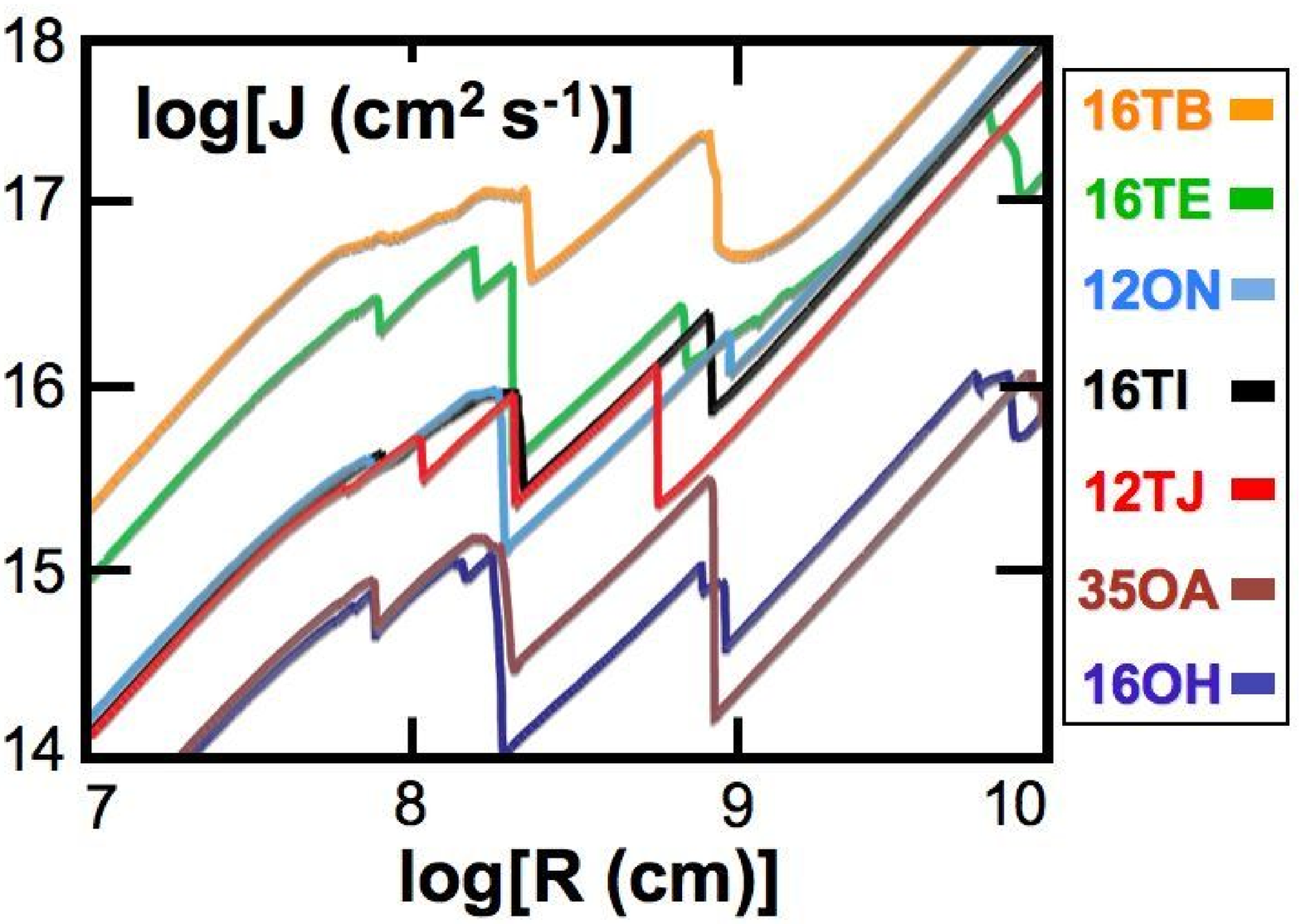} 
    \caption{Distributions of specific angular momentum versus radius
      for representative pre-SN evolutionary models, taken from
      WH06. The initial rise at low radii is followed by a sharp drop,
      secondary increase and fall, and outer increase into the stellar
      envelope. Note the form similarity and the relative
      normalization differences, which is due to assumed metallicity,
      mass loss and magnetic field effects (see
      Table~\ref{tab:inmodels} for model parameters).}
    \label{fig:Jdist}
  \end{center}
\end{figure}

The distribution of specific angular momentum for representative
models taken from WH06 is shown in Figure~\ref{fig:Jdist}. They
explored a large set of parameters, considering different initial zero
age main sequence masses, (12 - 35)$M_{\odot}$, various initial
rotation rates, (1 - 14)$r_{\rm g} $c in $J(R)$, initial
metallicities, (0.01 - 1)$Z_{\odot}$, mass loss, and the effects of
magnetic fields (see Table~\ref{tab:inmodels}). The resulting
distributions span slightly more than two orders of magnitude, and all
cases clearly show the same qualitative behavior. The general
background is that of a power law rise, segmented with sharp drops at
the interfaces of the various layers in the star ---for example, the
first drop corresponds to the boundary between the iron core and the
carbon-oxygen (CO) shell. The resulting layers, with varying degrees
of rotation and mass, must necessarily interact with one another as
they collapse onto the central object. As before, we chose model 16TI
as our fiducial angular momentum distribution, $J_{\rm 16TI}$.

\begin{table}\centering
\caption{Representative models from WH06.}
\label{tab:inmodels}
\begin{tabular}{cccccc}
\hline
\hline
Model & $M_{\rm ZAMS} [M_{\odot}]$ & $v_{\rm rot} [\mbox{km~s$^{-1}$}]$ 
& $Z [Z_{\odot}]$ & $M_{\rm loss} [M_{\odot}]$ & B-field \\
\hline
16TB & 16 & 305 & 0.01 & 0.71 & no \\
16TE & 16 & 305 & 0.01 & 4.02 & no \\
12ON & 12 & 400 & 0.1 & 1.07 & yes \\
16TI & 16 & 390 & 0.01 & 2.0 & no  \\
12TJ & 12 & 380 & 0.01 & 0.46 & yes \\
35OA & 35 & 380 & 0.1 & 0.6 & yes \\
16OH & 16 & 325 & 0.1 & 6.82 & yes  \\
\hline 
\end{tabular}
\end{table} 

We are unable to follow the collapse of the outer regions of the core
and envelope, covering the entire spatial and temporal range required
in the present simulations. This is mostly due to our choice of
location for the inner boundary, with which we are able to resolve the
transition from a QRI to a rotationally supported disk at a few
Schwarzschild radii. Nevertheless, the various time scale associated
with the arrival of particular shells at the centrifugal barrier can
be roughly estimated for these models as the free fall time, and the
previous analysis of high and low-angular momentum shells extended to
this regime, which we do in what follows. In reality, pressure support
within the envelope is non-negligible, so the corresponding times and
mass accretion rates are somewhat longer and lower, respectively (see
also the discussion by \citet{mw99} related to the latter). The
calculations by \citet{lindner09} show that the transitions in flow
morphology resulting from the structure of the progenitor can indeed
be manifested at late times.

\begin{figure}[h!]
  \begin{center}
  \epsscale{1.0}    
   \plotone{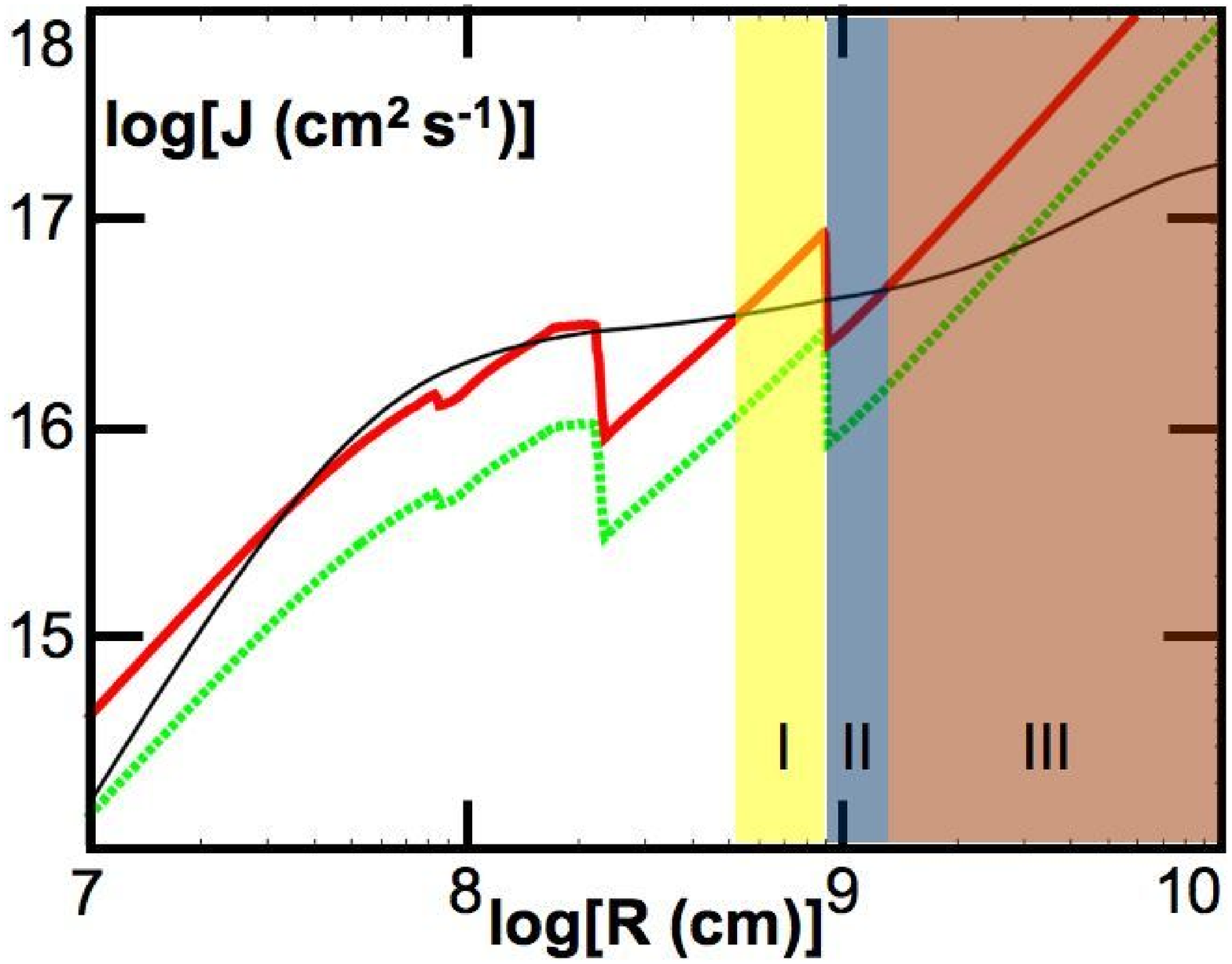} 
    \caption{Radial distribution of angular momentum for model 16TI of
      WH06, $J_{\rm 16TI}(R)$ (green dotted line) and critical curve
      for the formation of a centrifugal barrier $J_{\rm crit}$ (thin 
      black solid line) around the central Schwarzschild black hole
      with $M(R)$. The thick red line shows $ f \times J(R)_{{\rm
          16TI}}$ with $f=2$.}
    \label{fig:Jdist16TI}
  \end{center}
\end{figure}

In Figure~\ref{fig:Jdist16TI} we show $J(R)_{{\rm 16TI}}$ (green
dotted line) and $J_{\rm crit}(R)$ (black solid line), the critical
distribution for the formation of a centrifugal barrier assuming all
the matter at $r<R$ has been accreted by the black hole. It is clear
that up to $R^{*} \sim 2\times 10^{9}$cm, $J(R)_{{\rm 16TI}} \leq
J_{\rm crit}(R)$. Thus for this case, during the first ten seconds
(which corresponds to the free fall time required for the material
departing from $R^{*}$ to reach the black hole), we would have a QRI,
low-luminosity phase, as the one seen in the upper panel of
Figure~\ref{fig:pathways}. Given the uncertainties in the
normalization of $J(R)$ and the spread seen in Figure~\ref{fig:Jdist},
we now introduce a multiplicative scaling factor of order unity $f$
into $J(R)_{{\rm 16TI}}$ and compute the corresponding infall time
scales. In fact, $f$ may be tuned, for example so that the condition
$\mu=1/3$, discussed in \S~\ref{sec:states} is satisfied or not
between the initial disk and the first shell with low angular momentum
(red solid line in Figure~\ref{fig:Jdist16TI}, with $f=2$). Typically,
$f$ is of order unity, although re-normalizing some models to have
$\mu=1/3$ requires substantial deviations from this value (see
Table~\ref{tab:models}).

For a normalization leading to a distribution below that with
$\mu=1/3$, a centrifugally supported disk created from the material in
the shell in region ``I'' in Figure~\ref{fig:Jdist16TI} would be
destroyed by the QRI produced by the shell labeled ``II'', and finally
be followed by a second stable disk from region ``III''. Using the
free fall times for the interfaces between regions I, II and III we
find that after a one-second delay, an initially high-luminosity phase
of 1.1 seconds would be followed by a quiescent period of $\simeq
4.2$~s before activity resumed. Note that the initial black hole with
$1.6M_{\odot}$ is formed from material with $R \lesssim 2 \times
10^{8}$~cm (which includes the first large discontinuity in the
rotation rate), so no variable energy output emerges from this region.

\begin{figure}[h!]
  \begin{center}
  \epsscale{0.8}    
   \plotone{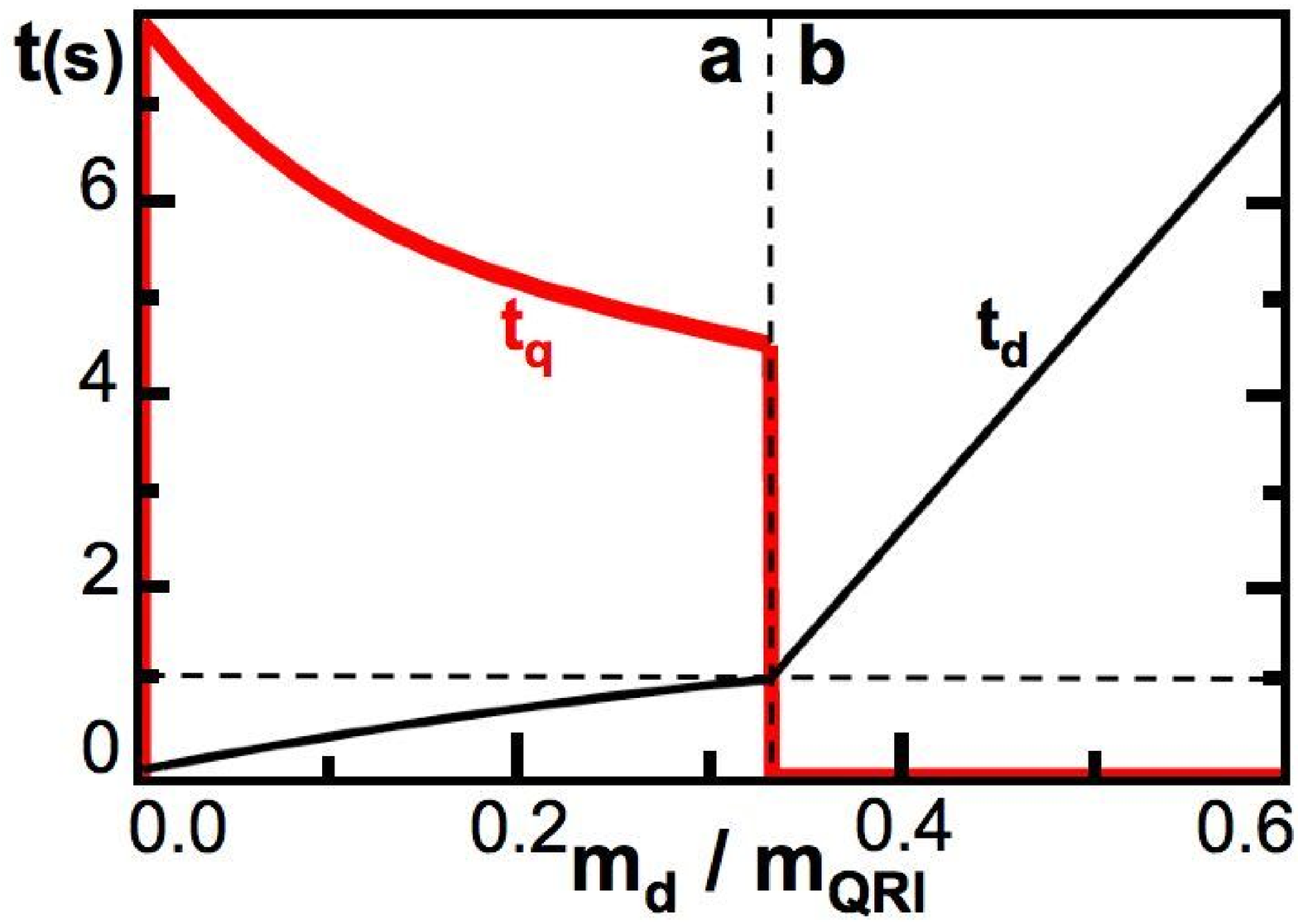} 
    \caption{Quiescent period, $t_{\rm q}$ (red line) and initial disk
      activity, $t_{\rm d}$ (black line) episodes as a function of the
      high angular momentum disk to low angular momentum shell mass
      ratio, $\mu=M_{\rm d}/M_{\rm sh}$. If $\mu \leq 1/3$, $t_{\rm
        q}+t_{\rm d}$ varies slowly, while for $\mu \ge 1/3$ the
      quiescent episode abruptly disappears and $t_{\rm q}=0$. The
      dashed horizontal line approximately indicates the jet crossing
      time, $t_{\rm cross}$, through the stellar envelope. In this
      case $t_{\rm q}$ is few seconds long, but can be significantly
      longer depending on the progenitor's rotation.}
    \label{fig:episodes}
  \end{center}
\end{figure}

For vanishing mass of a centrifugally supported disk ($\mu \rightarrow
0$), the luminosity is only that of the QRI at a level $L_{\nu} \simeq
10^{51}$~erg~s$^{-1}$\citep{lrr06,lclrr09}. As long as $0 < \mu <
1/3$ is satisfied, as in the case shown in Figure~\ref{fig:pathways}a, the low-angular momentum shell has greater mass than the centrifugally supported disk and a first high-luminosity transient followed by a quiescent period gradually appears, as seen in
Figure~\ref{fig:episodes}. We note that their respective durations,
$t_{\rm d}$ and $t_{\rm q}$ are {\em independent} of the assumed
strength of viscous transport, as they are purely dynamical
features. The initial episode lasts until the interface of the high
and low-angular momentum material, at the edge of the CO shell,
reaches the centrifugal barrier, $t_{\rm d} \simeq t_{\rm ff}(r_{\rm
  CO})$. The end of the subsequent quiescent interval, $t_{\rm q}$,
corresponds to the infall time of the second shell of rapidly rotating
gas (above the threshold for centrifugal support). The net result is that the sum between $t_{\rm d}$ and $t_{\rm q}$ varies slightly as a function of $\mu$ for a given initial mass configuration (Figure~\ref{fig:episodes}).
The neutrino luminosity in the high state is $L_{\nu} \simeq 10^{52}$~erg~s$^{-1}$. On the other hand, when the threshold value $\mu =1/3$ is reached, the quiescent interval vanishes as abruptly as the associated flow morphology. For larger values,
$L_{\nu}$ can in principle remain uninterruptedly at the high level as
long as there is mass feeding the disk. While for the particular case
shown here the quiescent period lasts for a few seconds, depending on
the initial conditions in the progenitor it can in principle be
substantially longer, matching those observed in GRB events.

\section{Discussion and prospects for observability}\label{sec:discussion}

We wish to stress that while we have presented the neutrino luminosity
as a measure of energy output, it is by no means the only one
possible, and should be viewed here as a proxy for central engine
activity, like the mass accretion rate with which it is very closely
correlated. One could equally use $\dot{M}$ (as long as it occurs in
state where energy is dissipated efficiently), or the power output in
magnetic outflows as a measure of the ability to drive relativistic
outflows. Our numerical scheme is geared towards appropriate handling
of thermodynamics and the associated neutrino emission, so it is
natural to rely on these properties when making quantitative
statements.

It is one thing to have a variation in the energy release close to the
BH, and another to have it manifest itself in the light curve of a
LGRB. Now, the minimum time required for activity to lead to high
energy emission is about 2~s, the crossing time for a relativistic
outflow originating in the inner regions to pierce the stellar
envelope \citep[see,
  e.g.,][]{mw99,aloy00,mwh01,aloy02,ramirezruiz02,zhang03}. The
crossing time is estimated here simply as $t_{\rm cross}=R_{\rm
  env}/(c/ \sqrt 3)$. In Table~\ref{tab:models}, we present a summary
of our results for the models shown in Figure~\ref{fig:Jdist} and
Table~\ref{tab:inmodels}, computed at the threshold where $\mu=1/3$
and a quiescent period is marginally present.

After an initial delay ($t_{\rm delay}$) between the onset of core
collapse and the initial rise in neutrino luminosity, a first disk is
active for an interval $t_{\rm d}$, followed by quiescence and late
disk creation. Only models with $t_{\rm d} \geq t_{\rm cross}$, in
boldface in Table~\ref{tab:models}, are potentially observable as
producing a quiescent period in the high-energy light curve under this
scheme. Note that $f$ is of order unity, but was scaled both to larger
and smaller values in order to find a condition at the threshold value, $\mu=1/3$.

\begin{table}\centering
\caption{Model evolution time scales. Events with quiescent periods.}
\label{tab:models}
\begin{tabular}{cccccc}
\hline
\hline
Model & $t_{\rm delay}[s] \ \tablenotemark{a} $ & $t_{\rm d}[s]$ & $t_{\rm q}[s]$ & $t_{\rm cross}[s]$ & $f$ \\
\hline
{\bf 16TB} & 0.5 & 2.3 & 9.6 &  1.8 & 0.6 \\
16TE & 1.3 & 0.9 & 3.0 &  1.8 & 2.0\\
12ON & 3.5& 0.6 & 1.6 &  1.5 & 1.9\\
{\bf 16TI}  &  2.3 & 1.1 & 4.2 &  1.0 & 2.0\\
12TJ  & 0.7 & 0.7 & 4.1 &  1.9 & 3.9 \\
{\bf 35OA} & 0.5 & 2.4 & 12.6 &  1.5 & 80\\
{\bf 16OH} & 0.4 & 2.1 & 1.4 &  2.0 & 39\\
\hline 
\tablenotetext{a}{Delay between onset of collapse and initial rise in $L_{\nu}$.}
\end{tabular}
\end{table} 

The most obvious limitation of the current study is the range of time
scales directly modeled. Nevertheless, we believe the current analysis
to be of relevance to the generic behavior in collapsing cores. Within
this range, the current calculations clearly show that the
characteristics of the energy output are closely correlated to the
distribution of specific angular momentum in the progenitor star. The
state transitions are abrupt, with the luminosity rising or dropping
by more than one order of magnitude, and reflect the naturally short
(ms) time scales in the vicinity of the accretor. The scaling to
larger radii and longer time scales can lead to quiescent period such
as those studied by \citet{ramirezruiz01a}, which in this case would
be related to dormant periods in the central engine
\citep{ramirezruiz01b}. \citet{ramirezruiz01a} found that a small
fraction, $\simeq$~15\%, of long GRBs exhibit at least one quiescent
interval, with about one quarter of these showing two such
episodes. Furthermore, the durations of the quiescent and subsequent
period of activity were found to be directly correlated. We believe it
is possible to account generically for this behavior under the present
picture in at least two relevant aspects.

First, the number of shell-like jumps in the distribution of angular
momentum within the star is small (one or two), accounting roughly for
the number of transitions observed. The second point is related to the
correlation itself. A short quiescent interval arises when the
low-angular momentum shell has a narrow radial extent. Its inner limit
at $r_1$ is fixed by the transition at the edge of the CO core and
remains approximately at the same radius, independently of the
normalization of angular momentum through the parameter $f$. The outer
boundary at $r_2$ on the other hand, varies with $f$ (see
Figure~\ref{fig:Jdist16TI}). Due to the latter and to the fact that
$J_{\rm crit}$ is a monotonically increasing function of radius,
higher overall angular momentum will lead to: (i) a short quiescent
interval, and (ii) a smaller circularization radius for the bulk of
the innermost matter within the centrifugally supported disk, $r_{\rm
  circ}[J(r_2)]$. Conversely, when the angular momentum normalization
is low, the quiescence period will be longer, and the circularization
radius for the majority of the innermost material will be greater. If
the evolution of the late--time disk is driven by the viscous
transport of angular momentum, the corresponding viscous time scale
and the duration of the associated accretion episode will scale
essentially with the initial disk radius. Thus, a longer quiescent
period will be followed by a longer period of disk activity. The
crucial point is that a lower global angular momentum normalization
will place the bulk of the initial accreting matter capable of forming
a centrifugally supported disk at a larger circularization
radius. There is in addition an effective upper limit for this radius
if the gas is to release its binding energy effectively, because if it
is too large, the density and temperature will not be sufficient for
neutrino cooling to operate.

As a final point, we may add that just as not all LGRBs exhibit this
behavior, likewise it is clear from this work that not all progenitors
are capable of producing the state transitions presented here. In
fact, \citet{pm10} have recently argued that the lack of flaring
activity in most LGRBs on long time scales is a signature of complete
mixing in the progenitor, which is fully in agreement with the line of
argument presented here. Once the deposition of energy in the inner
regions of the star has produced a relativistic jet capable of
traversing the stellar envelope, further variability may be reproduced
in the overall light curve \citep{zhang03}. Determining whether this
can power precursor activity is another matter, requiring the initial
episode of accretion to create a low density polar funnel in the star,
which remains to be addressed.

\acknowledgments We thank S.E. Woosley and A. Heger for making their
pre-SN models available. Part of this work was carried out during
visits to the University of California, Santa Cruz whose hospitality
is gratefully acknowledged. This work was supported in part by
CONACyT-83254 (DLC, WL), DGAPA-UNAM-IN-113007 (DLC, WL), the David and Lucile Packard Foundation (ER) and UCMEXUS (ER and WL). DLC
acknowledges support through a CONACyT graduate scholarship and a
DGAPA-UNAM postdoctoral fellowship. We thank the referee for constructive criticism which helped improve the original version of the manuscript.

\end{document}